\documentclass[rmp,floatfix,endfloats*]{revtex4}

\usepackage{natbib}
\usepackage{times,setspace}

\usepackage{amsmath, graphicx, amssymb,tabls}

\newcommand{\beq}{\begin{eqnarray}}
\newcommand{\eeq}{\end{eqnarray}}
\newcommand{\beqn}{\begin{equation}}
\newcommand{\eeqn}{\end{equation}}

\newcommand{\expt}[1]{\langle #1 \rangle }

\begin{document}

\title{Multiple paternity: determining the minimum number of sires of a large brood}

\author{A. Eriksson$^{a,b}$, B. Mehlig$^{a}$, M. Panova$^{c}$, C. Andr\'e$^{c}$, and K. Johannesson$^{c}$\\
$^a$\emph{\small Department of Physics, G\"oteborg University, G\"oteborg, Sweden}\\
$^b$\emph{\small Department of Marine Ecology, University of Gothenburg, SE-43005 G\"oteborg, Sweden}\\
$^c$\emph{\small Department of Marine Ecology-Tj\"a{}rn\"o, University of Gothenburg, SE-452 96 Str\"omstad, Sweden.}}

\begin{abstract}We describe an efficient algorithm for determining exactly the minimum number of sires consistent with the multi-locus ge
notypes of a mother and her progeny.
We consider cases where a simple exhaustive search through all possible sets
of sires  is impossible in practice (because it would take too long to complete).
Our algorithm for solving this combinatorial optimisation problem avoids visiting large parts
of search space which would not improve the  solution found so far
(i.e., result in a solution with fewer number of sires).
This is of particular importance when the number of allelic types in the progeny array is large
and when the minimum number of sires is expected to be large. Precisely in such cases it is important
to know the minimum number of sires: this number gives an exact bound on the
most likely number of sires estimated by a random search algorithm in a parameter region
where it may be difficult to determine whether it has converged.
We apply our algorithm to data from the marine snail, {\em Littorina saxatilis}.\\[0.2cm]
\end{abstract}

\maketitle

\section{Introduction}

A number of species from different taxa are known to mate numerous times during a mating season. Females of such species are likely to give birth to offspring fathered by more than one sire, and in some cases, the offspring may
be fathered by a large number of sires. For example, queens of the honeybee 
are known to leave the nest followed by hundreds of males \citep{Wat03}. 
Females of the saltwater fly mate many times a day during the mating season 
\citep{Bly06} and in species of periwinkles (marine gastropods) females mate repeatedly during the mating season which is several months long \citep{Sau90,Rei96}.

The degree of multiple paternity can be inferred  from
empirical data, obtained for example by genotyping females and offspring using high-resolution genetic markers, such as
 microsatellites or single nucleotide polymorphisms. 
To estimate the number of sires corresponding to a given multi-locus data set is usually straightforward 
when the number of offspring, the number of their allelic types, and the number of sires to be determined are small.

In the examples mentioned above, however, the levels of multiple paternity are often so high that the mathematical analysis
of the empirical data becomes very complex and time-consuming. In this paper we describe a new efficient algorithm to analyse
multiple paternity in empirical data sets when the paternal genotypes are unknown.
We apply the algorithm to data sets from the periwinkle species \emph{Littorina saxatilis}
exhibiting high levels of multiple paternity 
[it has been shown previously that at least 7-10 males are fathers to the offspring of one brood \citep{Mak07}].

Consider an example. Tab.~\ref{tab: 1} shows the
multi-locus genotypes of $42$ progeny  from a brood of a female periwinkle.
\begin{table*}
\setstretch{0.5}
\center
\raisebox{2cm}{%
\begin{tabular}{rllllllllll}
\hline
  \small\# & \small L1        & \small L2           & \small L3           & \small L4           & \small L5\\
\hline
 \small  0 & \{ \small  151 , 192 \}&\{              \small 227 , 242 \}&\{ \small  225 , 231 \}&\{ \small  217 , 223 \}&\{ \small  199 , 202 \}\\\hline
 \small  1 & \{ \small  151 , 204 \}&\{              \small 236 , 242 \}&\{ \small  222 , 231 \}&\{ \small  217 , 223 \}&\{ \small  199 , 202 \}\\
 \small  2 & \{ \small  151 , 211 \}&\{              \small 227 , 242 \}&\{ \small  210 , 231 \}&\{ \small  217 , 223 \}&\{ \small  199 , 223 \}\\
 \small  3 & \{ \small  151 , 195 \}&\{              \small 227 , 236 \}&\{ \small  210 , 225 \}&\{ \small  217 , 217 \}&\{ \small  196 , 202 \}\\
 \small  4 & \{ \small  151 , 181 \}&\{              \small 236 , 242 \}&\{ \small  213 , 225 \}&\{ \small  223 , 223 \}&\{ \small  199 , 199 \}\\
 \small  5 & \{ \small  151 , 195 \}&\{              \small 227 , 230 \}&\{ \small  210 , 225 \}&\{ \small  217 , 217 \}&\{ \small  196 , 202 \}\\
 \small  6 & \{ \small  151 , 192 \}&\{              \small 227 , 236 \}&\{ \small  222 , 225 \}&\{ \small  217 , 217 \}&\{ \small  184 , 199 \}\\
 \small  7 & \{ \small  192 , 208 \}&\{              \small 227 , 236 \}&\{ \small  219 , 231 \}&\{ \small  217 , 223 \}&\{ \small  193 , 199 \}\\
 \small  8 & \{ \small  151 , 173 \}&\{              \small 227 , 227 \}&\{ \small  213 , 231 \}&\{ \small  217 , 232 \}&\{ \small  193 , 199 \}\\
 \small  9 & \{ \small  151 , 208 \}&\{              \small 236 , 242 \}&\{ \small  219 , 231 \}&\{ \small  217 , 223 \}&\{ \small  199 , 199 \}\\
 \small  10 & \{ \small  151 , 151 \}&\{              \small 227 , 230 \}&\{ \small  216 , 231 \}&\{ \small  217 , 223 \}&\{ \small  184 , 199 \}\\
 \small  11 & \{ \small  192 , 208 \}&\{              \small 227 , 239 \}&\{ \small  225 , 225 \}&\{ \small  184 , 223 \}&\{ \small  199 , 223 \}\\
 \small  12 & \{ \small  173 , 192 \}&\{              \small 227 , 236 \}&\{ \small  213 , 231 \}&\{ \small  223 , 226 \}&\{ \small  193 , 199 \}\\
 \small  13 & \{ \small  151 , 201 \}&\{              \small 236 , 242 \}&\{ \small  216 , 231 \}&\{ \small  217 , 223 \}&\{ \small  199 , 202 \}\\
 \small  14 & \{ \small  151 , 201 \}&\{              \small 236 , 242 \}&\{ \small  213 , 231 \}&\{ \small  184 , 223 \}&\{ \small  202 , 223 \}\\
 \small  15 & \{ \small  169 , 192 \}&\{              \small 227 , 227 \}&\{ \small  219 , 231 \}&\{ \small  217 , 220 \}&\{ \small  202 , 205 \}\\
 \small  16 & \{ \small  192 , 201 \}&\{              \small 236 , 242 \}&\{ \small  219 , 225 \}&\{ \small  223 , 223 \}&\{ \small  199 , 199 \}\\
 \small  17 & \{ \small  151 , 192 \}&\{              \small 236 , 242 \}&\{ \small  213 , 225 \}&\{ \small  223 , 223 \}&\{ \small  193 , 199 \}\\
 \small  18 & \{ \small  192 , 201 \}&\{              \small 227 , 236 \}&\{ \small  219 , 231 \}&\{ \small  217 , 235 \}&\{ \small  202 , 202 \}\\
 \small  19 & \{ \small  151 , 173 \}&\{              \small 227 , 242 \}&\{ \small  225 , 231 \}&\{ \small  217 , 226 \}&\{ \small  193 , 199 \}\\
 \small  20 & \{ \small  173 , 192 \}&\{              \small 236 , 242 \}&\{ \small  213 , 231 \}&\{ \small  184 , 223 \}&\{ \small  202 , 202 \}\\
 \small  21 & \{ \small  182 , 192 \}&\{              \small 227 , 236 \}&\{ \small  213 , 225 \}&\{ \small  217 , 223 \}&\{ \small  190 , 202 \}\\
 \small  22 & \{ \small  151 , 204 \}&\{              \small 236 , 242 \}&\{ \small  222 , 225 \}&\{ \small  184 , 217 \}&\{ \small  202 , 202 \}\\
 \small  23 & \{ \small  192 , 204 \}&\{              \small 227 , 236 \}&\{ \small  222 , 225 \}&\{ \small  217 , 223 \}&\{ \small  199 , 199 \}\\
 \small  24 & \{ \small  151 , 181 \}&\{              \small 227 , 236 \}&\{ \small  219 , 231 \}&\{ \small  217 , 217 \}&\{ \small  199 , 202 \}\\
 \small  25 & \{ \small  192 , 201 \}&\{              \small 227 , 236 \}&\{ \small  213 , 231 \}&\{ \small  223 , 223 \}&\{ \small  199 , 199 \}\\
 \small  26 & \{ \small  151 , 208 \}&\{              \small 236 , 242 \}&\{ \small  225 , 231 \}&\{ \small  217 , 226 \}&\{ \small  199 , 202 \}\\
 \small  27 & \{ \small  151 , 195 \}&\{              \small 236 , 242 \}&\{ \small  210 , 225 \}&\{ \small  214 , 217 \}&\{ \small  184 , 199 \}\\
 \small  28 & \{ \small  151 , 208 \}&\{              \small 239 , 242 \}&\{ \small  225 , 231 \}&\{ \small  184 , 223 \}&\{ \small  199 , 199 \}\\
 \small  29 & \{ \small  151 , 201 \}&\{              \small 227 , 242 \}&\{ \small  210 , 231 \}&\{ \small  217 , 223 \}&\{ \small  202 , 223 \}\\
 \small  30 & \{ \small  192 , 217 \}&\{              \small 227 , 236 \}&\{ \small  216 , 231 \}&\{ \small  214 , 223 \}&\{ \small  199 , 202 \}\\
 \small  31 & \{ \small  151 , 195 \}&\{              \small 236 , 242 \}&\{ \small  210 , 225 \}&\{ \small  217 , 223 \}&\{ \small  184 , 202 \}\\
 \small  32 & \{ \small  192 , 206 \}&\{              \small 236 , 242 \}&\{ \small  219 , 225 \}&\{ \small  169 , 217 \}&\{ \small  196 , 202 \}\\
 \small  33 & \{ \small  151 , 208 \}&\{              \small 236 , 242 \}&\{ \small  219 , 231 \}&\{ \small  217 , 223 \}&\{ \small  199 , 205 \}\\
 \small  34 & \{ \small  151 , 204 \}&\{              \small 227 , 239 \}&\{ \small  210 , 225 \}&\{ \small  217 , 217 \}&\{ \small  196 , 199 \}\\
 \small  35 & \{ \small  192 , 194 \}&\{              \small 230 , 242 \}&\{ \small  219 , 225 \}&\{ \small  217 , 217 \}&\{ \small  202 , 202 \}\\
 \small  36 & \{ \small  192 , 192 \}&\{              \small 236 , 242 \}&\{ \small  222 , 231 \}&\{ \small  217 , 223 \}&\{ \small  199 , 199 \}\\
 \small  37 & \{ \small  151 , 194 \}&\{              \small 227 , 230 \}&\{ \small  210 , 225 \}&\{ \small  202 , 217 \}&\{ \small  202 , 202 \}\\
 \small  38 & \{ \small  151 , 206 \}&\{              \small 227 , 230 \}&\{ \small  219 , 225 \}&\{ \small  217 , 217 \}&\{ \small  202 , 205 \}\\
 \small  39 & \{ \small  151 , 168 \}&\{              \small 227 , 236 \}&\{ \small  222 , 225 \}&\{ \small  217 , 220 \}&\{ \small  199 , 205 \}\\
 \small  40 & \{ \small  192 , 195 \}&\{              \small 227 , 236 \}&\{ \small  216 , 225 \}&\{ \small  217 , 223 \}&\{ \small  196 , 202 \}\\
 \small  41 & \{ \small  192 , 192 \}&\{              \small 227 , 236 \}&\{ \small  213 , 231 \}&\{ \small  184 , 217 \}&\{ \small  199 , 202 \}\\
 \small  42 & \{ \small  192 , 217 \}&\{              \small 236 , 242 \}&\{ \small  222 , 225 \}&\{ \small  220 , 223 \}&\{ \small  199 , 199 \}\\
\hline
\end{tabular}}
\caption{\label{tab: 1}
{\em L. saxatilis}.
Multi-locus genotypes (five loci) of a mother (index $0$) and $n=42$ progeny from a clutch 
\citep[in preparation]{Bos09}.
}
\end{table*}
How does one determine
the number of sires from  the data
shown in Tab.~\ref{tab: 1}? More precisely, this question may be posed in at least three different ways:
what is the {\em actual} number of sires, as opposed to the {\em most likely} number,
or the {\em minimum} number of sires consistent with the mother and her progeny array?

Unless one is able to directly observe the matings it is in general not possible 
to determine the actual number of sires for an array such as the one shown in Tab.~\ref{tab: 1}.
An alternative is to instead determine the most likely number of sires consistent
with the multi-locus genotypes of mother and progeny \citep{Wan04}. In this approach it
is commonly assumed that the population is in Hardy-Weinberg equilibrium, that all loci
are subject to neutral evolution, and that all loci are in pairwise linkage equilibrium.
The most likely set of sires is determined by a random search algorithm \citep{PressTeuk},
locally maximising the likelihood constrained by requiring consistency with mother and progeny.
A random search however does not guarantee convergence. The only way to make sure that
the algorithm has converged is to compare with an exhaustive search.

Exhaustive search algorithms have been published in the literature. An example available for download is GERUD 
\citep{Jon01,Jon05}. The exhaustive search is commonly
conducted as follows: a list of paternal alleles at all loci is determined from the 
alleles in the progeny array, subtracting those of the mother. If for a given child at a given
locus the paternal alleles cannot be uniquely extracted, both are kept in the list. From this
list, a set of potential sires   is obtained by constructing all possible multi-locus
paternal haplotypes. This list is pruned by removing individuals which are not consistent with 
any of the progeny in the data. The minimum number of sires   is determined by 
exhaustively searching this pruned set: first the algorithm tests whether a single
father from this set is consistent with the progeny array. If this is not the case,
all pairs of sires   are tested, if necessary all triplets of sires, and so forth.
This algorithm ensures that the minimum number of sires consistent with the data
are found. 
This algorithm has been successfully used in many circumstances. It works well
when the list of paternal alleles is not too long, and when the minimum number of fathers
to be determined is not too large. 

The minimum number of sires for the data sets shown in Tab.~\ref{tab: 1}
is found to be twelve (using the algorithm described below). With the search algorithm
summarised above, one would have to go through 
a prohibitively large number of sets of possible sires. Since the number of sets to search
typically increases combinatorially with the number of sires, the exhaustive search
in the algorithm proposed by \cite{Jon05} is limited to maximally six sires for
a given sample. In practice, 
when the number of sires is five or six, the algorithm
may take very long to converge \cite{Sef08}.  In many practical applications \cite{Ama08,Van08,Sim08,Tak08,Ris08,Son07,Por07} the number of sires determined with GERUD
does not exceed four. (It should be
noted that GERUD 2.0 allows the user to truncate the pruned set of sires in
an {\em ad-hoc} fashion in order to check for up to eight sires. But due
to the truncation, the minimum number of sires is potentially overestimated. By how much
is unclear.)

A maximum likelihood approach indicates that the number of sires
in samples of {\em L. saxatilis} \citep{Mak07} is typically larger than six.
We have therefore developed a new algorithm for determining the minimum number
of sires consistent with a given progeny array, such as that shown in Tab.~\ref{tab: 1}.
The algorithm is described in Sec. \ref{sec: methods}. 
It makes it possible to exactly determine the minimum number of sires
for the empirical data sets listed in Tab.~\ref{tab: 2}, in seven of the nine
data sets the minimum number of sires is found to be larger than six.
\begin{table}[t]
\center
\begin{tabular}{ccccc}
\hline
Female             & n   & GERUD 2.0 & COLONY  &  MinFathers \\
\hline
Tab.~\ref{tab: 1}$^\ddagger$     & 42    & -$^\ast$    &13$^\$$  & 12 \\
E1-M$^\dagger$        & 21    & -         & 8          & 7 \\
E7-M$^\dagger$        & 22    & -         & 9          & 8 \\
E3-M$^\dagger$        & 21    & -         & 9          & 8 \\
E4-M$^\dagger$        & 21    & -         & 7          & 7 \\
S1-M$^\dagger$        & 23    & 4         & 4          & 4 \\
S5-M$^\dagger$        & 23    & -         & 9          & 8 \\
S2-M$^\dagger$        & 23    & 5         & 5          & 5 \\
S3-M$^\dagger$        & 23    & -         & 10         & 8 \\
\hline     
\end{tabular}
\caption{\label{tab: 2}
Comparison of the results (minimum or most likely
number of sires, respectively) of three algorithms
for different progeny arrays. $n$ is the number of offspring in the sample. \lq MinFathers' denotes
the results of the algorithm described in section \ref{sec: methods}, in all cases
all five loci are taken into account.
Footnotes:
$^\ast$The symbol $-$ denotes that the number of sires exceeded the maximum possible number for an exhaustive search in GERUD 2.0.
$^\$$ Most likely number of fathers according to COLONY.
$^\dagger$Data taken from \citep{Mak07}.
$^\ddagger$Data taken from \citep[in preparation]{Bos09}.
}
\end{table}
Further, the new algorithm enables us to estimate the number of sires
for a data set when convergence of maximum-likelihood algorithms
is difficult to ascertain. Fig.~\ref{fig: 1} for example shows a run of the maximum-likelihood
algorithm COLONY \citep{Wan04} (which estimates the most likely number of 
fathers given the population allele frequencies) for the data shown in Tab.~\ref{tab: 1}, compared to  
the minimum number of sires (which is twelve for this data set). 
Fig.~\ref{fig: 2} raises the question how likely it is that the 
minimum number of sires  is equal to the actual number of sires
for a given sample. Using our algorithm we have investigated this question
employing data sets generated by a coalescent algorithm within a
step-wise mutation model: the answer depends upon the properties
of the population in question.  In cases where the probability that
the two numbers are the same is high, we can infer that
the actual nmber of sires in this situation can be reliably estimated
from empirical samples (it should be emphasized that the result
of our algorithm always is an exact lower bound).
Last but not least, we employ our algorithm to answer the following
question: given an empirical data set, how could one most efficiently increase 
the accuracy of the estimate of the number of sires: by genotyping more loci for a given set
of progeny, or by genotyping more progeny for a given number of loci?
Again, the answer depends upon the properties of the population in question.

The remainder of this article is organised as follows. In Sec. \ref{sec: methods} we describe
the new algorithm, called \lq MinFathers'.  We also briefly describe how we produced
artificial samples using the coalescent in order to test the new algorithm.
Our results are summarised in Sec. \ref{sec: results}. Conclusions are drawn
in Sec.~\ref{sec:discussion}.

\section{Methods}
\label{sec: methods}

\subsection{An efficient search algorithm}
In this section we describe our new algorithm which determines the minimum
number of sires consistent with a given progeny array, such as that
shown in Tab.~\ref{tab: 1}.

To find the minimal set of sires is equivalent to finding a partition of the progeny,
such that all progeny in a given member of the partition can be inherited from a single father. 
A father is represented by a list of the two alleles at each locus. Each paternal allele may either have a definite value, or no value may yet have been assigned to this allele.
Each progeny too is represented by a list of alleles, one for each locus when the allele
of the mother has been subtracted.
Usually, the allelic types are uniquely determined. There are, however, two exceptions:
when a genotype error has occurred, and 
when the mother and the offspring are identical and heterozygous. In this case, at the locus in question,
there are two possible alleles for the progeny. 
We ignore these complications for the moment, and return to them later, after having described the algorithm in its simplest form.

In our algorithm, the most general common father for a set of progeny is found through a sequence of merging operations. 
This operation maps the two most general fathers of two sets of progeny to
the most general father of the combined set. Since we are always searching for minimum number of fathers, the fathers are  always taken to
be heterozygous at each locus (or some loci remain undetermined).  

The merging of two fathers $f$ and $f'$ proceeds independently
at each locus (since free recombination is assumed). 
Assume that a common father $f$ for a set of $j$ progeny has been found. Now add another progeny
to this set.  Assume that the new progeny has allelic type $a$ at a given locus. Its most general
father has the genetic configuration $\{a,0\}$ at this locus. The most general
common father $f'$ of the joint set of $j+1$ progeny is obtained by merging $f$ and $\{a,0\}$ as
described in Tab.~\ref{tab:merge}. 
At a given locus, 
the father $f'$ may have several possible configurations, depending on the configurations of $f$ and the father of $p$. 
In the table, the asterisk denotes that the corresponding allelic type has not yet been determined, or is unknown.
The most general father of a single progeny is 
$\{a,\ast\}$.

\begin{table}[t]
\center
\begin{tabular}{ccl}
\hline
father $f$ & new father& common father $f'$ \\
\hline
 any                    & \{ $\ast$, $\ast$\}  & any \\
  \{$\ast$ , $\ast$\}   & \{ $a$, $\ast$\}     & \{$a$, $\ast$\} \\
  \{$a$, $\ast$\}       & \{ $a$, $\ast$\}     & \{$a$, $\ast$\} \\
  \{$a$, $\ast$\}       & \{ $b$, $\ast$\}     & \{$a$, $b$\} \\
  \{$a$, $b$\}          & \{ $a$, $\ast$\}     & \{$a$, $b$\} \\
  \{$a$, $b$\}          & \{ $b$, $\ast$\}     & \{$a$, $b$\} \\
  \{$a$, $b$\}          & \{ $c$ , $\ast$\}    & no common father \\
\hline
\end{tabular}
\caption{\label{tab:merge}
The possible outcomes of merging  a father $f$ with a new father, resulting in the common father $f'$. 
Undetermined alleles are represented by an asterisk ($\ast$), and $a$, $b$, and $c$ are three different allelic types. 
In the first row, the allele of the new progeny and thus the configuration of its father is undetermined, so that $f' = f$. 
In the final row, $f'$ 
would have to contain three different alleles, which is impossible. Therefore, no common father exist in this case.
}
\end{table}

Now consider a partition $F$ of $j$ progeny. We introduce the following terminology. 
A partition is called \emph{valid} if for each element of the partition there is a 
common father for all progeny in the element. In other words, a valid partition of progeny corresponds
to a set of fathers for the progeny in question.
Our algorithm can now 
be formulated as follows: for each valid partition $F$ of progeny $1, \ldots, j$, generate all valid partitions $F'$ 
of progeny $1, \ldots, j+1$ by adding a new progeny $j+1$ to each element of $F$ provided
the new father merges with the common father of the element of $F$ into
a valid common father. 
Starting from an empty set of fathers, $F = \emptyset$, and a set of progeny $S$, 
we can find all valid sets of fathers of $S$ by this algorithm.

\begin{figure*}[t]
\centerline{\includegraphics[width=522pt]{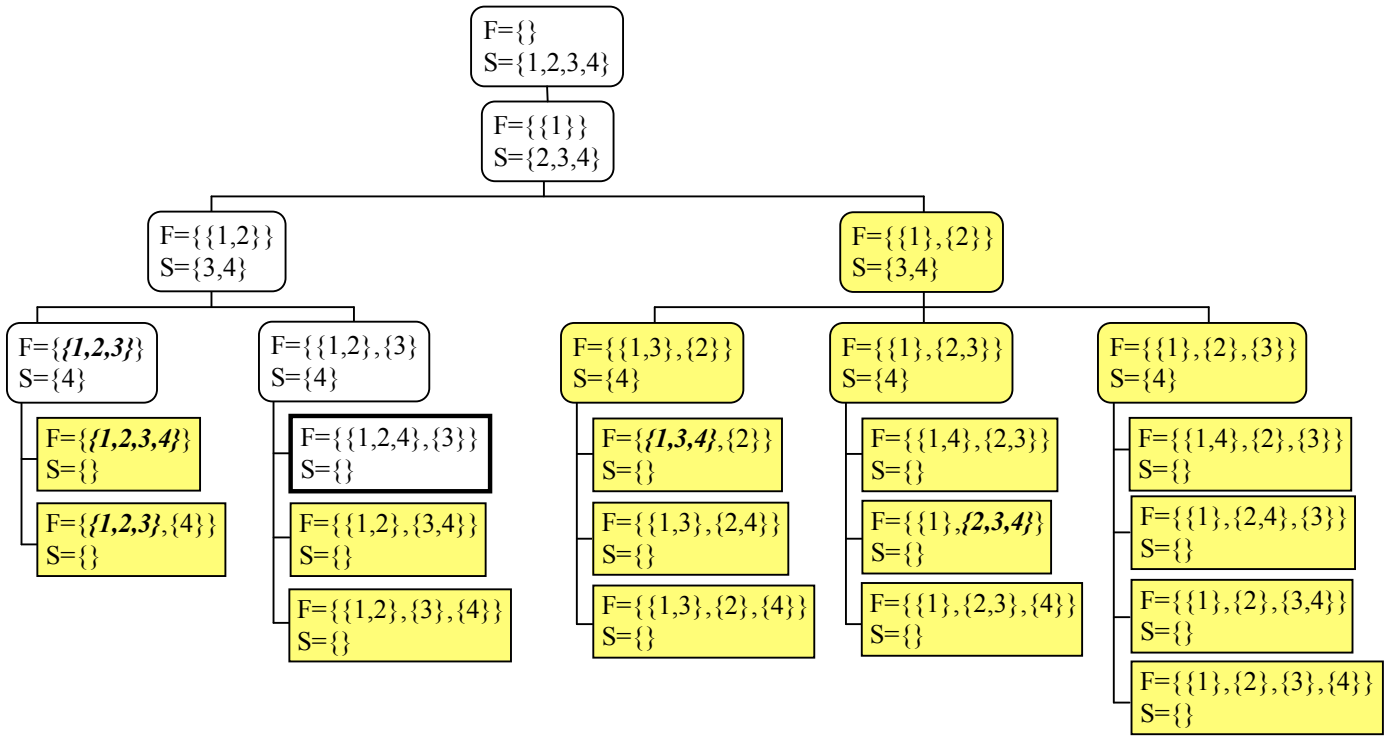}}
\caption{\label{fig: 2}
Shows a search tree for four progeny constructed by the algorithm
described in Sec. \ref{sec: methods}.  Each progeny has two loci, and at each locus the allele corresponding to the mother has been subtracted, so
that each progeny is described by a list of two alleles (one for each locus):  $p_1 = (a,b)$, $p_2 = (c,d)$, $p_3 = (c,e)$, and $p_3 = (a,d)$.
It is assumed that $a$, $b$, $c$, and $d$ are four different allelic types.
The figure illustrates how large parts of the search tree can be cut away (yellow) because they need not be visited. This may considerably speedup the algorithm. 
In the figure, each box represents the state of the algorithm at each iteration; $F$ is the set of fathers (each father is represented by the list of offspring assigned to it), and $S$ is the set of offspring the algorithm has yet to be assigned to a father. The states are visited from top to bottom, and from left to right, following the lines that emanate from the bottom of each node, except the terminal nodes (shown as squares). The algorithm stops when the set $S$ is empty. A set of individuals that cannot inherit from a single father is shown in bold italic font (i.e. the father is invalid). The states which are coloured yellow are never visited, either because they descend from a state with an invalid father, or because it can be seen that the state does not lead to a solution with fewer fathers than the best state found so far in the search.
}
\end{figure*}

It is possible to find the minimum number of fathers from this method by taking the minimum modulus of all partitions found. This is usually much better than first generating the full set of partitions of the progeny, and subsequently 
checking which partitions are valid; how much more efficient this
is depends on the data. Our algorithm for finding the minimum number of fathers is summarised in Fig.~\ref{alg:minfathers}. It recursively builds all valid partitions of a set of progeny $S$, except some partitions that can be shown to not correspond to the minimum number of fathers. In Fig.~\ref{fig: 2} we show a search tree for a simple set of 
four progeny. Each progeny has two loci, and at each locus the allele corresponding to the mother has been subtracted, so 
that each progeny is described by a list of two alleles (one for each locus):  $p_1 = (a,b)$, $p_2 = (c,d)$, $p_3 = (c,e)$, and $p_3 = (a,d)$. 
It is assumed that $a$, $b$, $c$, and $d$ are four different allelic types.

\begin{figure}
\centerline{\includegraphics[width=252pt]{f1.epsi}}
\caption{\label{alg:minfathers} Algorithm for finding the minimum number of fathers. $S$ is the set of progeny, $F$ is the set of fathers ($F = \emptyset$ initially), $\hat{n}$ is the minimum number of fathers for the whole set of progeny found so far 
($\lceil|S|/2\rceil$ initially). See the text for an explanation of the algorithm. Under which circumstances a
loop over variants $q$ of $p$ is necessary is explained in the text.
}
\end{figure}

We conclude this section by  emphasizing four important points.
First, note that if we have a given number of fathers for progeny $1,\ldots,j$, the number of fathers for progeny $1,\ldots,j+1$ 
must be at least as high. Hence, whenever the set of fathers is at least as large as the minimum number found so far, we can stop searching for partitions of $S$ based on the current partition. When we have a complete partition of $S$ which is smaller than the minimum found so far, we can update the minimum. As a starting point, we can use any valid upper bound; in the present implementation we use the trivial bound that the minimum number of fathers is  $n^* \le \lceil|S|/2\rceil$. Using our algorithm, we have
\begin{equation}
 n^* = \text{MinFathers}(\emptyset, S, \lceil|S|/2\rceil).
\end{equation}
%
%

Second, the algorithm       as described above is valid only if the genetic material inherited from the father at each locus is uniquely determined. In general, this is not the case, as pointed out above.
Instead, there may be one or more loci with two possible choices. In this case, when generating the valid partitions containing a given progeny we loop over all possible variants of alleles in these loci. In practical data, it is rare to have more than one such locus, but if many loci are considered this could cause  problems: the number of variants of the allele that may need to be considered is $2^m$, where $m$ is the number of loci with multiple choices in the progeny. For data sets where this is a problem, it is possible to extend the algorithm to a more complex merging operation, where for each locus of a father, we keep track of all possible pairs of alleles that can simultaneously match all progeny deriving from the father.

Third, if we find that some of the progeny not yet included in the current set of fathers can be directly inherited from any of the fathers (i.e. the father contains the necessary genetic material at all loci), we can safely remove them from from the set of progeny. 
In other words, if a new progeny $p$ can be directly inherited from a father $f$, a merge $f'$ between $f$ and $p$ will lead to $f'=f$, i.e. no change in $f$. Hence, it is clear that given the present set of fathers, it is not possible to find another way of merging these progeny which will lead to fewer fathers for the whole set. In our algorithm, whenever we add  a new father to the partition, we remove the progeny that can be directly inherited from the new father, before recursively searching for new partitions.

Fourth, in general the key to an efficient solution of a combinatorial problem such as the present one lies in cutting away as large parts of the search space as early as possible. In the present context, this means that if we can consider the most constraining progeny first, we can discard partitions that will not be valid for the whole set of progeny, or that will be larger than the minimum, at an early stage. We sort the progeny first with respect to the number of undetermined  
symbols (if any), and where then number of no-care symbols is equal, with respect to the number of multiple choice loci in the progeny, so that individuals with multiple-choice or undetermined loci will be considered first. 
Before the sorting, we check the multiple choice loci. Consider a multiple choice locus in a given progeny. If the two alleles only occur together at that locus in all progeny, or if only one of the loci occur alone or together with some third allele, we can safely replace the multiple choice by the most frequent of the allelic types. Picking the least frequent allelic type can only exclude some possible merges that would have been possible with the more common allele. However, if both alleles occur also alone or with some third allelic type, we cannot safely conclude which choice will lead to the minimum number of fathers. In this case it is necessary to keep both options.

\subsection{Generating samples using the coalescent}
\label{sec: coalescent}
In this section we describe how we have used the coalescent to generate artificial samples
in order to test the search algorithm describe in the previous section.

In order to understand under which circumstances the minimum number of fathers is equal to the true number of fathers, 
we generate $n_{\rm p}$ progeny with a known number of fathers, $n_{\rm f}$, as follows. First, the gene genealogies of the $L$ loci in a mother and $n_{\rm f}$ fathers are generated according to the standard neutral coalescent theory for an unstructured population with constant size $N$. Because we assume that the loci are unlinked, the gene genealogies of different loci are statistically independent. In each branch of the genealogies, mutations occur with probability $\mu$ per generation, so that the number of mutations in a branch of $T$ generations is Poisson distributed with mean $\mu T$. In the coalescent, time is measured in units of $2N$ generations and the mutation rate is given by the scaled parameter $\theta = 4N\mu$.

We model microsatellite data using the stepwise mutation model, where each mutation leads to either the gain or the loss of a single repeat unit \citep{kim75:dis,kim78:ste}. Thus, given the genealogy, for each locus we start from the most recent common ancestor of the whole sample, and assign it allele $0$ (in the stepwise mutation number, only differences in the number of repeat units are relevant). We then recursively generate the alleles of each node in the genealogy by generating the stepwise mutations along each branch as described above, until we have assigned the alleles for all individuals in the sample. 

Given the allelic types of the mother and the fathers, we produce the offspring as follows: for each progeny we pick a randomly chosen father, such that each father is equally likely to be picked. If in the end not all fathers have been picked for at least some offspring, we repeat the whole
process until this is the case. This guarantees that the true number of fathers is exactly $n_{\rm f}$. For each locus in the progeny we then form the progeny according to Mendelian inheritance from the mother and the father, by picking one allele from the mother and one from the father. This procedure guarantees that the marginal distribution of the number of offspring per father is approximately binomial, which is consistent with the neutral theory and with empirical data for the snails \citep{Mak07}.

The performance of \lq MinFathers' depends on the amount of genetic variation (determined by the mutation rate $\theta$) and on the number of progeny per father, $n_{\rm p/f} = n_{\rm p}/n_{\rm f}$. When $\theta$ is small, the minimum number of fathers is small and the algorithm terminates quickly. Also when $\theta$ is large the algorithm is efficient because it can usually eliminate impossible merges at an early stage. When $\theta$ is intermediate and $n_{\rm p/f}$ is large, however, the algorithm may have to investigate a significant fraction of the possible combinations, and in these cases it may not be practical to use the algorithm. Despite this caveat, Tab.~\ref{tab: 2} and the results present in the next section show that the algorithm can be used on empirical data with a large number of progeny and across a large range of parameters for $\theta$ and $n_{\rm p/f}$.

\section{Results}
\label{sec: results}
In this section we describe the results obtained with the new algorithm proposed
in section \ref{sec: methods}. 

\subsection{Application to empirical data}
\begin{figure}
\includegraphics[width=252pt]{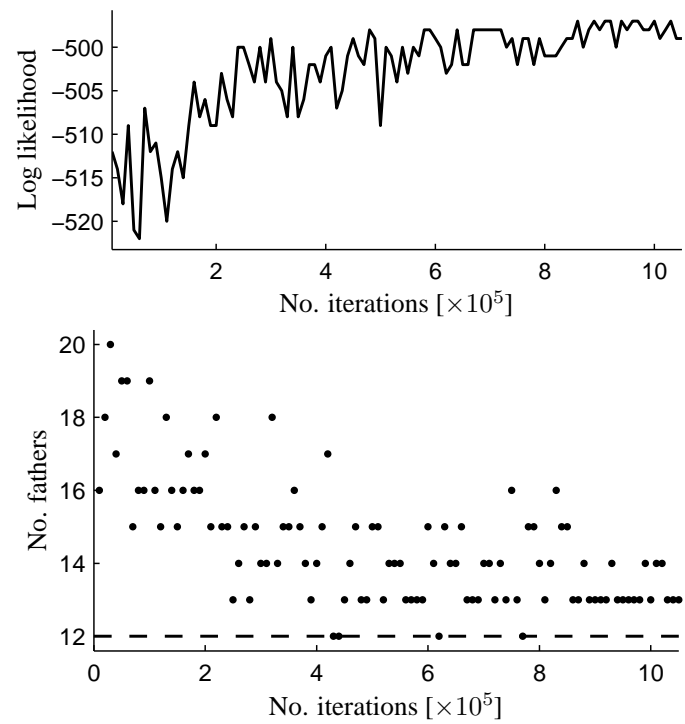}
\caption{\label{fig: 1}
Shows a run of COLONY \citep{Wan04} for
the data shown in Tab. \ref{tab: 1}. Top: The time evolution of the log likelihood of the data.
Bottom: The most likely number of sires as a function of the number of iterations (dots). Also shown is
the exact lower bound provided by the new algorithm, MinFathers (dashed line).}
\end{figure}

We have applied our algorithm to the {\em L. saxatilis} data by \cite{Mak07}, and to a
new {\em L. saxatilis} data set \citep[in preparation]{Bos09}, given in Tab.~\ref{tab: 1}.
We begin by describing our results for the new data set (Tab.~\ref{tab: 1}). 
The original data contains more progeny than listed in Tab.~\ref{tab: 1}. 
Using our algorithm we find that the minimum number of sires is twelve, as given in the first row of Tab. \ref{tab: 2}.
The algorithm GERUD 2.0 could not be run
because this algorithm determines the exact solution only for up to six sires. 
We have also run COLONY \citep{Wan04} (which estimates the most likely number of sires), 
and the corresponding results are shown in Fig.~\ref{fig: 1}. It is 
not entirely clear whether the algorithm has converged; the log likelihood may have reached a plateau but it is not clear whether further exploration of the state-space may yield still higher likelihood values. It is therefore valuable to have the exact lower bound
(also shown in Fig.~\ref{fig: 1}) from our new algorithm.

Using the new algorithm, MinFathers, we have re-analysed the data of \cite{Mak07};
the corresponding results are also given in Tab. \ref{tab: 2}. The last eight data sets in Tab.~\ref{tab: 2}
were analysed using GERUD by \cite{Mak07} and are broadly
consistent with the corresponding results of COLONY. It must be noted however
that in \cite{Mak07},  the search was performed on three
loci only, and using an {\em ad hoc} truncation of the possible set of fathers.
It is therefore of interest to determine what the minimum number of sires
actually is. The corresponding results are shown in Tab. \ref{tab: 2}, 
and provide an exact lower bound for the most likely number of
fathers determined by \cite{Mak07}. Except in two cases, GERUD 2.0
could not be run because the true minimum number of fathers exceeded the
maximum value of six.

The results summarised in Tab. \ref{tab: 2} raise the question of
how much larger than the exact minimum one expects the most
likely number of fathers to be. The answer depends upon
the population model, and upon the parameters describing it,
such as the mutation rate $\theta$, the number $L$ of loci,
and the number $n$ of progeny in the data.  This question
is addressed in the following section.

\subsection{The difference between the minimum and the most likely number of sires}
In this section we consider a population in Hardy-Weinberg equilibrium, we assume
that all loci are subject to neutral evolution, and that all loci are in pairwise linkage equilibrium.
We pose the question: how much larger than the minimum number is the most likely
number of sires in a brood of a given mother? To this end we generated
samples using the coalescent as described in Sec. \ref{sec: coalescent}.

\begin{figure}[t]
\centerline{\includegraphics[width=252pt]{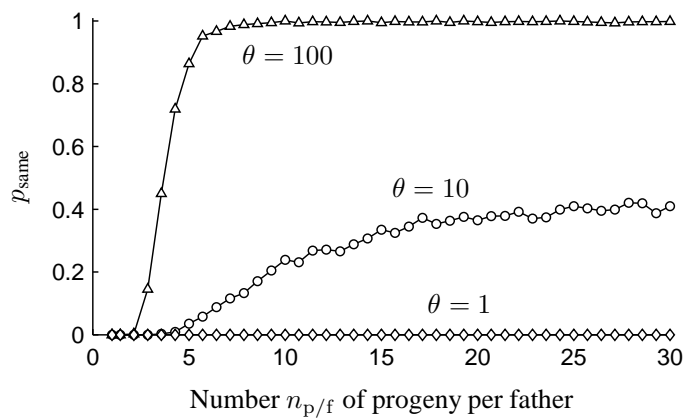}}
\caption{\label{fig:p_corr_npf}
The probability $p_\text{same}$ that the number of fathers $n_{\rm f}$ is equal to the minimum number of fathers, as a function of the number of progeny per father, for three values of $\theta$; $\theta = 1$ (diamonds), $\theta = 10$ (circles), and $\theta = 100$ (triangles). 
The actual number $n_{\rm f}$ of fathers is $7$, and there are three loci ($L = 3$). Each data point is based on $1000$ families.
}
\end{figure}

\begin{figure}[t]
\centerline{\includegraphics[width=252pt]{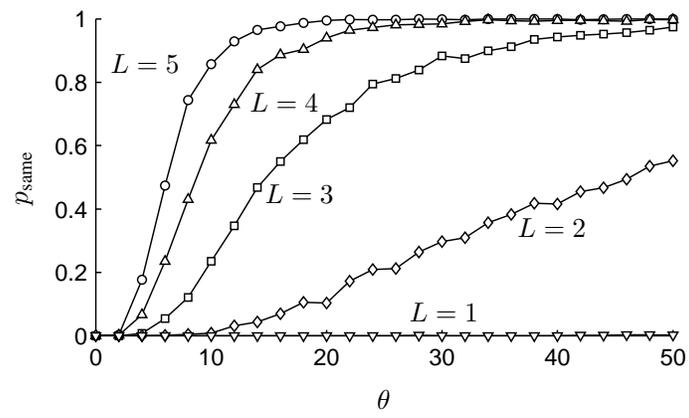}}
\caption{\label{fig: 3} 
Shows the probability that the most likely number of fathers is larger than the minimum number as a function of the mutation rate $\theta$ for five different values of the number of loci $L$. Each data point is based on $1000$ families with $n_\text{f} = 7$ fathers and $n_\text{p} = 70$ offspring ($n_\text{p/f} = 10$).
}
\end{figure}

Figs.~\ref{fig:p_corr_npf} and \ref{fig: 3} summarise our results. 
First, Fig.~\ref{fig:p_corr_npf} shows the probability that the number of fathers is equal to the minimum number of fathers, $p_\text{same}$, as a function of the number $n_{\rm p/f}$ of progeny per father, for three values of the
mutation rate $\theta$; $\theta = 1$, $\theta = 10$, and $\theta = 100$. When $\theta$ is large, we observe a sharp transition where $p_\text{same}$ increases from almost zero to almost unity. 
When $\theta$ is small, however, sampling many offspring does not result in a signficiant increase of $p_\text{same}$ because the fathers are too genetically similar. For intermediate values of $\theta$, $p_\text{same}$ does increase with the number of progeny per father, but never reaches unity because the fathers are still significantly genetically correlated.

Second, Fig.~\ref{fig: 3} shows how $p_\text{same}$ changes as a function of $\theta$ for five different values of the number of loci $L$. 
When $L=1$, the minimum number of fathers is always smaller than the true number of fathers (unless there is only one father),
therefore $p_{\rm same}$ equals zero. For $L \ge 2$ we find that $p_\text{same}$ increases as a function of both $\theta$ and $L$. The extent to which the minimum number of fathers agrees with the true number of fathers depends on the probability that the fathers are all genetically distinct. When $L$ is increasing, this happens for smaller values of $\theta$, as indicated by the increasingly sharp transitions from $p_\text{same} = 0$ to $p_\text{same} = 1$ in the figure.

Thus, when the mutation rate, number of loci, and number of offspring per father are sufficiently high
so that $p_{\rm same}\approx 1$, the minimum number of fathers almost always equals the true number of fathers. 
As a consequence, the resulting number of sires contributing to a given family does not depend on whether
 the population is structured or panmitic. In other words, the result is insensitive to assumptions about the underlying population structure.

\section{Discussion}
\label{sec:discussion}

We conclude with a discussion of, first, how the minimum number of fathers is influenced by genotyping errors. Second, we
address the following question. If the aim is to increase the probability of deducing the true number of fathers from empirical data
such as Tab.~\ref{tab: 1}, is it better of to sample more offspring or more loci?
 
Microsatellites can be prone to genotyping errors \citep[see reviews by][]{Hoffman:2005, DeWoody:2006, Selkoe:2006}. Sources of error when genotyping microsatellites include stuttering (appearance of PCR products one or more repeats shorter than an actual allele), allele dropout (non-amplification of one of the two alleles in a heterozygote, usually a longer one), non-specific PCR-products due to annealing of primers to multiple sites, null alleles (alleles with a mutation in the primer region, which prevent their amplification in PCR) and, finally, mistyping and other mistakes during manual scoring of the results.

Changes to the number of repeats (PCR stuttering events) may cause the minimum
number of fathers to appear larger than it actually is.
PCR stuttering events in a microsatellite locus are usually incremental (single additions or deletions of a repeat unit). 
Hence, when the sample size is large, it is likely that the resulting alleles are present in other fathers (the effect of PCR stuttering is similar to mutations occurring during meiosis). 
Therefore, when the frequency of stuttering events is small, 
the effect on the minimum number of fathers is expected to be small. 

When only a single allele is amplified (e.g. because of allele dropout or null alleles), the minimum number of fathers of the sample 
may in- or decrease.
If the frequency of such errors is small, however, 
the effect is expected to be small: assuming that the incidence of these errors are independent across different loci, the likelihood that the errors change the minimum number of fathers decreases rapidly with increasing number of loci.

We now turn to the question of how to best increase the accuracy of estimating the true number of fathers.
When the number of offspring is large, as in the marine snails, there are two possibilities for increasing the probability that we can deduce the true number of fathers. One may either sample the same loci in more offspring, or one may sample more loci in the offspring we already have. Which is the better option? Our results show that increasing the number of loci generally provides the quickest way of increasing the probability of finding the true number of fathers from the minimum number of fathers, but whether this is feasible or not depends on the availability and cost of additional high-quality markers 
(see Figs.~\ref{fig:p_corr_npf} and \ref{fig: 3}). It may be less costly to sample more individuals with fewer loci, if possible. In this case, however, the accuracy may be limited by the number of offspring available but also the genetic variation in the loci. If few loci are sampled, and the mutation rate is low, our results show that it may not be possible to increase the accuracy by sampling more individuals beyond a certain limit, which is determined by the probability that the fathers share the same alleles.

In order to relate the theoretical discussion
of Sec.~3.2 to the empirical data (Tabs.~\ref{tab: 1} and \ref{tab: 2}), we have estimated the parameter $\theta$ from the data in \citet{Mak07} using two standard estimators, $\hat\theta_v = \expt{(x_i - y_j)^2}$ \citep{Wehrhahn:1975} and $\hat\theta_F = (F^{-2} - 1)/2$ \citep{Ota:1973}. 
Here $x_i$ and $y_j$ are alleles of progeny $i$ and $j$ from mothers $x$ and $y$, respectively, and $F$ is the homozygosity (the probability that $x_i = y_j$). It is known that $\hat\theta_v$ is unbiased [but has a large variance \citep{Zhivotovsky:1995}], whereas $\hat\theta_F$ is biased for large values of $\theta$, but has a smaller variance. For the data in Tab.~\ref{tab: 2} we obtain  $\hat\theta_v = 124$ and $\hat\theta_F = 49$ when averaged over all five loci. While these estimates are uncertain because of the small sample size, we see from Figs.~\ref{fig:p_corr_npf} and \ref{fig: 3} that the number of progeny sampled in \citep{Mak07} ($n_p = 21$) is probably too low to reliably estimate the true number of fathers (the estimated $n_{p/f}$ is the range $2\ldots 4$ for these data). Increasing the number of loci is not likely to help very much. The progeny sampled were chosen from large families (of 70 to 100 progeny). Our results indicate that in this case, the best strategy for increasing the accuracy of the number of fathers is to sample still more progeny. 
Indeed, analysis of the full set of progeny this data set indicates that the true number of fathers in these families is significantly higher than the minimum number of fathers reported in Tab.~\ref{tab: 2} 
\citep[in preparation]{Bos09}.

{\em Acknowledgments}. 
We thank J. Bostr\"om, T. Hofving, T. Areskoug and T. M\"akinen for providing data, 
and the Swedish Research Council, the Linneus initiative Adaptation to Changing Marine Environments (ACME) and 
the Centre for Theoretical Biology at Gothenburg University for support. 
\bibliographystyle{elsarticle-harv}

\end{document}